\documentclass[aps,a4paper,longbibliography,twocolumn,superscriptaddress]{revtex4-2}
\usepackage{times}
\usepackage{comment}
\usepackage{graphicx}
\usepackage{feynmf}
\usepackage{tabularx}
\usepackage{amsmath}
\usepackage{amstext}
\usepackage{amssymb}
\usepackage{xfrac}
\usepackage{amsfonts}
\usepackage{longtable,booktabs}
\usepackage{url}
\usepackage{subfigure}
\usepackage{dsfont}
\usepackage{amsbsy}
\usepackage{dcolumn}
\usepackage{amsthm}
\usepackage{bm}
\usepackage{esint}
\usepackage{multirow}
\usepackage{mathrsfs}
\usepackage{extarrows}
\usepackage{datetime}
\usepackage{epsfig}
\usepackage{xcolor}
\usepackage{setspace}
\usepackage{mathtools}
\usepackage{physics}
\usepackage{cases}
\usepackage{slashed}
\usepackage{pifont}
\usepackage[super]{nth}

\usepackage[
colorlinks=true,
citecolor=blue,
linkcolor=blue,
urlcolor=magenta,
filecolor=magenta
]{hyperref}

\usepackage{cleveref}

\newcommand{\assign}{:=}

\newcommand{\tmop}[1]{\ensuremath{\operatorname{#1}}}

\begin{document}
\title{Macroscopic Particle Transport in Dissipative Long-Range Bosonic Systems}

\author{Hongchao Li}
\email{lhc@cat.phys.s.u-tokyo.ac.jp}
\affiliation{Department of Physics, University of Tokyo, 7-3-1 Hongo, Tokyo 113-0033, Japan}

\author{Cheng Shang}
\email{cheng.shang@riken.jp}
\affiliation{Analytical Quantum Complexity RIKEN Hakubi Research Team, RIKEN Center for Quantum Computing (RQC), Wako, Saitama 351-0198, Japan}

\author{Tomotaka Kuwahara}
\email{tomotaka.kuwahara@riken.jp}
\affiliation{Analytical Quantum Complexity RIKEN Hakubi Research Team, RIKEN Center for Quantum Computing (RQC), Wako, Saitama 351-0198, Japan}
\affiliation{RIKEN Cluster for Pioneering Research (CPR), Wako, Saitama 351-0198, Japan}
\affiliation{PRESTO, Japan Science and Technology (JST), Kawaguchi, Saitama 332-0012, Japan}

\author{Tan Van Vu}
\email{tan.vu@yukawa.kyoto-u.ac.jp}
\affiliation{Center for Gravitational Physics and Quantum Information, Yukawa Institute for Theoretical Physics, Kyoto University, Kitashirakawa Oiwakecho, Sakyo-ku, Kyoto 606-8502, Japan}

\date{\today}

\begin{abstract}
     Dissipation in quantum many-body systems provides a more general and experimentally realistic perspective on particle transport than closed quantum systems. In this work, we determine the maximal speed of macroscopic particle transport in dissipative bosonic systems featuring both long-range hopping and long-range interactions. By developing a generalized optimal transport theory for open quantum systems, we rigorously establish the relationship between the minimum transport time and the source-target distance, and investigate the maximal transportable distance of bosons. We demonstrate that optimal transport exhibits a fundamental distinction depending on whether the system experiences one-body loss or multi-body loss. Moreover, we present the minimal transport time and the maximal transport distance for systems with both gain and loss. We observe that even an arbitrarily small gain rate enables transport over long distances if the lattice gas is dilute. Importantly, we generally reveal that the emergence of decoherence-free subspaces facilitates the long-distance and perfect transport process. Additionally, we derive an upper bound for the probability of transporting a given number of particles during a fixed period in the presence of particle loss. Possible experimental protocols for observing our theoretical predictions are also discussed.
\end{abstract}
\maketitle
\section{Introduction}
The maximal velocity of information propagation is a central topic in quantum mechanics. Information causality, a fundamental constraint in relativistic quantum mechanics, gives rise to an upper bound for information propagation, known as a light cone.~Remarkably, Lieb and Robinson pointed out that quantum information propagation in locally interacting quantum spin systems on lattices also exhibits an effective light cone structure~\cite{Lieb1972,hastings2010localityquantumsystems}.~This result, named the Lieb-Robinson (LR) bound, demonstrates locality even in non-relativistic quantum many-body systems and has undergone substantial improvements and extensive extensions since its discovery. The LR bound has been extended to general interacting quantum spin systems and fermionic systems~\cite{Hastings2006long,Hauke2013,Cevolani_2016,Tran2019,ChenLucas:2019,Guo_Andrew2020,Colmenarez2020,KuwaharaPRL2021} and interacting bosonic systems~\cite{Kuwahara2021,Yin2022,Faupin_LR2022,Lemm2023info,Kuwahara2024}. Recently, it has been further generalized to open quantum systems~\cite{LRopen_2010,Barthel2012,Sweke_2019,Rahul2024,PhysRevLett.121.170601}. The LR bound has now become a powerful tool in condensed matter physics~\cite{Bachmann2017,Bravyi2006,Bravyi_order2010,Bruno2007,Hastings_LSM2004,Shenglong2024}, quantum information science~\cite{Hastings_2007,Landau2015,Nahum2018,Shenglong_NP2020,PhysRevB.111.184311,Hamazaki_2022}, high energy physics~\cite{Roberts2016,Maldecena2016,Kobrin2021}, etc.  

Despite the great success in applying the Lieb-Robinson bound, its extension to bosonic systems remains limited due to the fact that the norm of the bosonic operators is unbounded. Constant progress has been made in understanding the maximal speed for information propagation in bosonic systems with short-range hopping by restricting the initial boson number~\cite{Kuwahara2024,Faupin_LR2022} to one-dimensional systems~\cite{Yin2022} or to suitable initial states~\cite{Lemm2023info}. However, for general unbounded operators in high-dimensional systems, the traditional arguments yield only unsatisfactory results on the maximal information-propagation speed~\cite{Kuwahara2021}. On the other hand, some studies focus on the speed limit for macroscopic particle transport, which quantifies the minimum time required for a given number of particles to propagate to a reachable regime. Recent results have elucidated the maximal speed of bosonic macroscopic particle transport, from the Bose-Hubbard model~\cite{Faupin2022,Faupin_LR2022,Vu2023topo,Faupin2025} to more general systems with both long-range hopping and long-range interactions \cite{Vu_2024,lemm2023microscopic}.

Nevertheless, the existing findings are limited to closed quantum systems. Due to intrinsic chemical reactions and inelastic collisions between particles~\cite{Ni2008,Ni2010,Liu2020,Bause2021,Bause2023,stevenson2023ultracold,Tetsu2014,Molony2014,Sebastian2016,Sebastian2023,KimYeong2004,Kraemer_threebody2006,Schindewolf2022,Maxime2024,PhysRevA.110.013322,Tomita_2019,Li2024BEC}, photoassociation~\cite{Arthur_Photon2019,McKenzie2002,Junker2008,doi:10.1126/sciadv.1701513,Hongchao2023} or interactions with external environments~\cite{10.1093/acprof:oso/9780199213900.001.0001,Lindblad1976,10.1063/1.522979,shang2023,RevModPhys.88.021002,Rivas_2012,Weiss_2012,Shang2024}, these systems are usually not strictly closed and inevitably suffer from dissipation, such as dephasing and particle loss, which is common in atomic, molecular, and optical systems. The dissipation causes the breakdown of the optimal transport theory in closed quantum systems since the traditional optimal transport theory only applies to the transport between two balanced particle distributions with equal total particle numbers. However, particle loss disrupts the balance between the two distributions. While previous studies have investigated particle transport in Markovian open quantum systems~\cite{Breteaux_2022,Breteaux2024,sigal2025lightconeboundsmarkov}, they are primarily restricted to few-body regimes. In contrast, the maximal transport speed in general many-body open quantum systems remains largely unexplored.

In this work, we develop the macroscopic particle transport theory for generic bosonic systems with long-range interaction and long-range hopping in the presence of particle loss. Inspired by the methods in Ref.~\cite{Vu_2024}, we construct the optimal transport theory applicable to open quantum systems and demonstrate the relationship between the transport time and the distance between the source and target regions for the Lindbladian dynamics. Specifically, for local particle loss, we show that the optimal transport time for the case of one-body loss in Eq.~\eqref{eq:single} is even longer than that with multi-body loss in Eq.~\eqref{eq:many}, which can be attributed to the existence of decoherence-free subspaces~\cite{Lidar1998} of the systems. The emergence of decoherence-free subspaces inherently supports long-distance and perfect (i.e., 100\%) transport (see Result 2 below).

Furthermore, we analyze Lindbladian dynamics with both gain and loss and demonstrate that particle gain can effectively reduce transport time, as shown in Eq.~\eqref{eq:gain_loss}. We compare the maximal transport distance between the case with only one-body loss and the case with both gain and loss, as given in Eqs.~\eqref{eq: distance-1} and \eqref{eq: distance-2}. Our results indicate that particle gain extends the transport distance. In particular, if the lattice gas is dilute, even an infinitesimally small gain rate can enable arbitrarily long transport distances over the lattice. We also derive an upper bound for the probability of transporting a given number of bosons between regions in a fixed period, as shown in Eq.~\eqref{eq: closed_P}. We conclude by discussing potential experimental protocols to test the predicted transport bounds and highlighting open questions that arise from our study. 

\section{Main Results}
\subsection{Local particle loss}
 We consider a quantum many-body system on a $D$-dimensional lattice with $\Lambda$ standing for all sites. For an arbitrary subset $Z \subseteq \Lambda$, the number of sites in $Z$ is denoted by $\left| Z \right|$, while the system size is represented by $\left| \Lambda  \right|$. The bosonic Hamiltonian can be written as
\begin{equation}\label{close_Hamiltonian}
  H := \sum_{i \neq j \in \Lambda} J_{i j} (t) b_i^{\dagger} b_j + \sum_{Z \subseteq \Lambda} h_Z (t),
\end{equation}
where $b_i^\dag $ and ${b_i}$ are the bosonic creation and annihilation operators at the site $i \in \Lambda $, respectively. Without any assumptions on boson-boson interactions in Eq.~(\ref{close_Hamiltonian}), ${h_Z}\left( t \right)$ is an arbitrary function of the boson number operators $\{ \hat{n}_i \}_{i \in \Lambda}$, where ${\hat{n}_i}: = b_i^\dag {b_i}$. The hopping amplitudes ${J_{ij}}\left( t \right)$ are symmetric and upper bounded by a power law in the Euclidean distance, i.e., $\left| {{J_{ij}}(t)} \right| \le {{J \mathord{\left/
 {\vphantom {J {\left\| {i - j} \right\|}}} \right.
 \kern-\nulldelimiterspace} {\left\| {i - j} \right\|}}^\alpha }$ for a positive constant $J$ and the power decay $\alpha $ satisfies $\alpha  > D$. In other words, both the long-range hopping and long-range interaction can be considered.
 
\begin{figure}
  \centering
  \includegraphics[scale = 0.42]{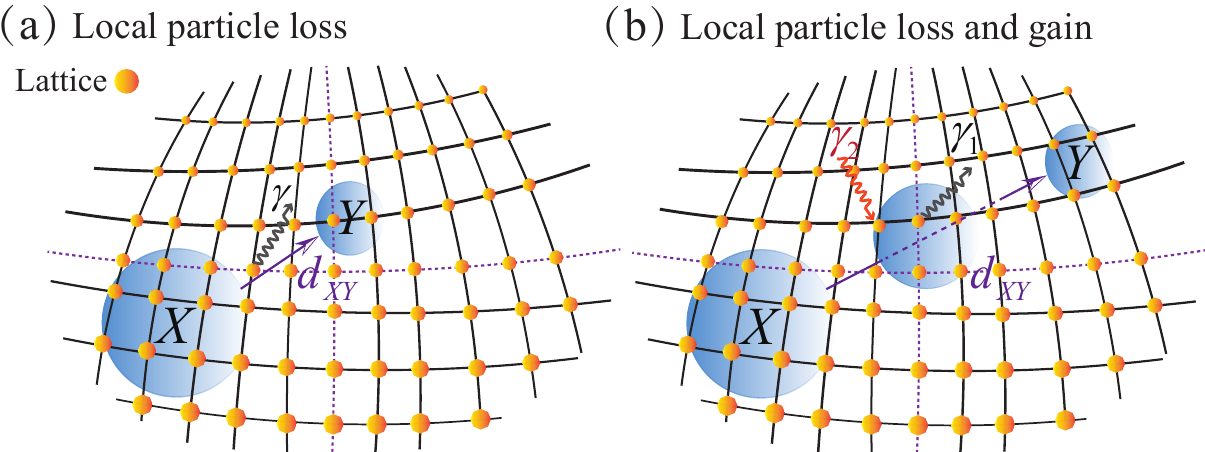}
  \caption{Schematic of particle transport in long-range bosonic open systems, wherein bosons are transported from region $X$ to a separate region $Y$ over a distance ${d_{XY}}$. $(\rm{a})$ The scenario with local particle loss includes Result 1 and Result 2. (b) The scenario corresponds to Result 3, where local particle loss (${\gamma _1}$) and gain (${\gamma _2}$) coexist. Particle gain counteracts loss, extending the maximal transport distance.}
  \label{Fig-setup}
\end{figure}

Let ${\rho _t}$ be the density matrix of the system. In the presence of dissipation, its dynamics follows the Lindblad equation:
\begin{equation}\label{eq: Lindblad}
  \frac{d \rho_t}{d t} = i [\rho_t,H] + \sum_{i} \frac{\gamma}{2}
  (2 L_i \rho_t L_i^{\dagger} - L_i^{\dagger} L_i \rho_t - \rho_t
  L_i^{\dagger} L_i),
\end{equation}
where we assume the Born-Markov approximation, which is valid in the weak-coupling regime. Here, $\gamma$ represents the jump rate, and $L_i$ denotes the Lindblad operator. In this study, we focus on the local Lindbladian, where the Lindblad operator $L_i$ only acts on one site. Specifically, we consider two types of Lindblad operators: one characterizing purely particle loss and the other incorporating both particle loss and gain~\cite{dephasing_operator}. The corresponding transport process is illustrated in Fig.~\ref{Fig-setup}. Figure~\ref{Fig-setup}(a) depicts the case of one-body loss or multi-body loss, while Fig.~\ref{Fig-setup}(b) presents the scenario where particle loss and gain coexist.

Consider a scenario where a fraction $\mu  \in \left( {0,1} \right]$ of all bosons is transported from region $X$ to a distant region $Y$, which in the open quantum systems is defined as
\begin{equation}\label{eq: def_trans}
    x_Y(\tau)-y_{X^c}(\tau)\geq\mu.
\end{equation}
Here $x_Y(\tau)N$ is the total particle number in region $Y$ at time $\tau$ with the Hamiltonian $H$ and $y_{X^c}(\tau)N$ is the total particle number in region $X^c:=\Lambda\setminus X$ at time $\tau$ when the Hamiltonian is switched off throughout the process, with $N$ being the initial total boson number on the lattice and $X\setminus Y$ representing the set of sites in region $X$ that is not included in region $Y$. We define the transport in this way because the change in particle number within region $Y$ can arise from three sources: particle exchange with the environment, transport from region $X^c\setminus Y$, and transport from region $X$. Our definition effectively eliminates contributions from the first two, isolating the net transfer from $X$ to $Y$. Accordingly, the inequality \eqref{eq: def_trans} indicates that $\mu N$ particles are effectively transported from $X$ to $Y$ within time $\tau$, taking into account both dissipation and hopping processes on the lattice. The distance between any two disjoint sets $X$ and $Y$ is defined as the minimum distance between their respective sites, i.e., $d_{XY} := \min_{i \in X, j \in Y} \|i - j\|$. The lattice satisfies $|i[r]\setminus i[r-1]|\leq\varphi r^{D-1}$, where $\varphi$ is an $O$(1) constant determined by the lattice structure and $ i[r]$ denotes the ball of radius $r$ centered at site $i$. For the above setting, we develop the macroscopic particle transport theory for long-range open systems by introducing a unified approach based on optimal transport theory. Specifically, we derive the following fundamental limits on the minimum time required for particle transport. The detailed proofs are outlined in the Methods section \ref{sec:method}.

\vspace{6pt}
\noindent \textbf{Result 1}. For the local one-body-loss case, i.e., $L_i=b_i$, the transport time $\tau$ satisfies
\begin{equation}\label{eq:single}
    \tau e^{- \gamma \tau} \geq\kappa_1^{\varepsilon} d_{X
  Y}^{\alpha_{\varepsilon}},
\end{equation}
where $\kappa_1^{\varepsilon}:=\mu/(J\varphi\zeta(\alpha-\alpha_{\varepsilon}-D+1))$ and $\zeta(x)$ represents the Riemann-zeta function. Here, $\alpha_{\varepsilon}:=\min(1,\alpha-D-\varepsilon)$ with $ 0 < \varepsilon < \alpha - D $ being an arbitrary number.

Compared with the extant result in closed quantum systems~\cite{Vu_2024}
\begin{equation}\label{eq: closed_tau}
    \tau\geq\kappa_1^{\varepsilon} d_{X
  Y}^{\alpha_{\varepsilon}},
\end{equation}
we find that the transport time $\tau$ is suppressed by an exponential decay, as the particle number decreases exponentially over time. This fact follows directly from the equation
\begin{align}
    \frac{d\langle \hat{N}\rangle}{dt}&=\frac{\gamma}{2}\sum_i\langle b_i^\dagger[\hat{N},b_i]+[b_i^\dagger,\hat{N}]b_i\rangle=-\gamma\langle \hat{N}\rangle,\nonumber
\end{align}
which yields the exponential decay $N_t=Ne^{-\gamma t}$. Here, $\hat{N}$ denotes the total particle number operator, and $N_t=\langle \hat{N}\rangle:=\tmop{tr}(\hat{N}\rho_t)$ is the expected total particle number at time $t$.

From Eq.~\eqref{eq:single}, we observe that there exists a maximal particle number that can be transported from region $X$ to region $Y$. The upper bound for $\mu$ takes the form of
\begin{equation}\label{eq: upper-mu}
    \mu\leq \min \left( \frac{J \varphi \zeta (\alpha - \alpha_{\varepsilon} -
  D + 1)}{e\gamma d_{X Y}^{\alpha_{\varepsilon}}}, 1 \right)
\end{equation}
by optimizing the transport time $\tau$ in Eq.~\eqref{eq:single}. If the one-body loss rate is significant, it is more likely that the amount $\mu N$ of bosons can never be transported from region $X$ to region $Y$ even in the infinite-time limit. Furthermore, we consider another transport limit of the bosons, defined as the farthest distance $d_l$ for which one boson can be transported. In the case of minimal transport (i.e., $\mu=1/N$), the transport limit is given by
\begin{equation}\label{eq: distance-1}
    d_l=\left(\frac{NJ\varphi\zeta(\alpha-\alpha_{\varepsilon}-D+1)}{e\gamma}\right)^{1/\alpha_{\varepsilon}}.
\end{equation}
This result indicates that if $N$ bosons are initially confined at the origin, their transport remains restricted within a region of radius $d_l$. This constraint, introduced by particle loss, highlights a form of locality and provides an upper bound on the effective trapping range in experiments. 

\vspace{6pt}
\noindent \textbf{Result 2}. For the local multi-body-loss case, i.e., $L_i=b_i^n$ with $n>1,n\in\mathbb{Z}$, the transport time $\tau$ satisfies
\begin{equation}\label{eq:many}
    \tau \geq\kappa_1^{\varepsilon} d_{X
  Y}^{\alpha_{\varepsilon}}.
\end{equation}

Notably, despite the presence of particle loss, this lower bound remains identical to that of the closed system [cf.~Eq.~\eqref{eq: closed_tau}]. The difference between the one-body-loss case and the multi-body-loss case stems from the presence of decoherence-free subspaces~\cite{Lidar1998}. For one-body loss, it is evident that the decoherence-free subspace consists only of the vacuum state. However, for multi-body loss, the decoherence-free subspace is given by the projection $\mathds{P}=\sum_{\textbf{n}}|\textbf{n}\rangle\langle \textbf{n}|$, where $|\textbf{n}\rangle$ is the Mott state with no more than ($n-1$) particles on each site. If the initial state is dilute and the dynamics remain confined to the subspace $\rm{Ran} \ \mathbb{P}$, particle transport proceeds without dissipation. As a result, the optimal transport under multi-body loss coincides with that of a closed quantum system. This explains why Eq.~\eqref{eq:many} is independent of the exponent $n$. A particularly natural example exhibiting the decoherence-free subspace is a system that contains at most ($n-1$) particles. However, the particle number is strictly limited in this example. For quantum many-body systems with $N\geq n$, the equality in Eq.~\eqref{eq:many} can still be asymptotically saturated by the following Hamiltonian:
\begin{equation}
  H = \sum_{i \neq j} J_{i j} (b_i^{\dagger} b_j+b_j^{\dagger}b_i)+ U \sum_i \hat{n}_i (\hat{n}_i - 1),
\end{equation}
where the limit $U\to\infty$ imposes the hard-core boson constraint, corresponding to the Mott-insulating phase in which only up to a single occupancy per site is allowed~\cite{Sengupta2005}. In this case, if the initial state lies within the decoherence-free subspace, the quantum state remains confined to the subspace due to the suppression of multi-particle occupation. Consequently, the transport process effectively mimics that of a closed quantum system. Therefore, this nontrivial decoherence-free subspace enables perfect transport over arbitrary distances without limitations on $\mu$ and $d_l$. For instance, particles initialized in a Mott state can be perfectly transported using the protocol proposed in Ref.~\cite{Vu_2024}. In contrast, for one-body loss, no nontrivial decoherence-free subspace exists, making the equality in Eq.~\eqref{eq:many} unattainable; instead, a tighter bound is provided in Eq.~\eqref{eq:single} of Result 1. Experimentally, multi-body loss typically arises from internal chemical reactions or photoassociation.

\subsection{Local particle loss and gain}
Thus far, we have only considered local particle loss. In what follows, we further investigate particle transport behavior in the presence of both particle loss and gain.

\vspace{6pt}
\noindent \textbf{Result 3}. For the Lindbladian with both local one-body loss and one-body gain, where the Lindblad operator takes the form of $L_i:=\sqrt{\gamma_1/\gamma}b_i+\sqrt{\gamma_2/\gamma}b_i^{\dagger}$ with $\gamma_2<\gamma_1$, the transport time satisfies
\begin{align}\label{eq:gain_loss}
    \tau e^{-\Delta\gamma\tau}+\frac{\gamma_2}{\Delta\gamma}\mathcal{N}^{-1}\left(\frac{1-e^{-\Delta\gamma\tau}}{\Delta\gamma}-\tau e^{-\Delta\gamma\tau}\right) \geq\kappa_1^{\varepsilon} d_{X Y}^{\alpha_{\varepsilon}},
\end{align}
where $\Delta\gamma:=\gamma_1-\gamma_2$, and $\mathcal{N}:=N/|\Lambda|$ is the initial particle density on the lattice. Here, $\gamma_1$ and $\gamma_2$ represent the loss and gain rates, respectively.

We note that Result 1 is a special case of Result 3 with $\gamma_1=\gamma$ and $\gamma_2=0$. Notably, the left-hand side of Eq.~\eqref{eq:gain_loss} does not vanish in the long-time limit, making it fundamentally different from the one-body-loss case. This indicates that gain intrinsically changes the particle transport process. 

From Eq.~\eqref{eq:gain_loss}, the maximal fraction of particles that can be transported from region $X$ to region $Y$ is given by
\begin{equation}
    \mu\leq\min\left(\frac{J\varphi\zeta(\alpha-\alpha_{\varepsilon}-D+1)}{d_{XY}^{\alpha_{\varepsilon}}}B(\tau_c),1\right),
\end{equation}
where
\begin{align}\label{eq:max.frac}
B(t):=\left\{
\begin{array}{rcl}
\frac{e^{-\Delta\gamma t}}{(\Delta\gamma)^2 t}+\frac{\gamma_2\mathcal{N}^{-1}}{(\Delta\gamma)^2} & \text{if} & {\mathcal{N}>\frac{\gamma_2}{\Delta\gamma},}\\
\frac{\gamma_2\mathcal{N}^{-1}}{(\Delta\gamma)^2} & \text{if} & {\mathcal{N}\leq\frac{\gamma_2}{\Delta\gamma},}
\end{array} \right.
\end{align}
and $\tau_c:=(\Delta\gamma-\gamma_2/\mathcal{N})^{-1}$. 
We examine two limiting cases of this result. In the dense limit ($\mathcal{N} \to \infty$), $B(\tau_c)$ asymptotically approaches $1 / (e\Delta\gamma)$, indicating an enhancement of the upper bound on $\mu$ compared to Eq.~\eqref{eq: upper-mu}. In contrast, in the dilute limit ($\mathcal{N} \to 0$), the upper bound diverges as $B(\tau_c) = \gamma_2 \mathcal{N}^{-1} / (\Delta\gamma)^2 \to \infty$.
This divergence suggests that an arbitrary number of particles can be transported from $X$ to $Y$ when the lattice is sufficiently large. The underlying reason for this behavior can be explained as follows. The dark state $|\psi\rangle$, which satisfies $L_i|\psi\rangle=0$, takes the product form of squeezed states as
\begin{equation}\label{eq: dark}
    |\psi\rangle=\bigotimes_{i}\left(\exp\left[\frac{r}{2}(b_i^2-b_i^{\dagger 2})\right]|0\rangle_i\right),
\end{equation}
where $r=\rm{arctanh}(\sqrt{\gamma_2/\gamma_1})$. This state is also the unique steady state of the Lindbladian when the Hamiltonian is switched off. 
One can explicitly show that the average particle number per site in $|\psi\rangle$ is given by $\langle\psi|\hat{n}_i|\psi\rangle=\gamma_2/\Delta\gamma:=\bar{n}$, which coincides with the critical particle density in Eq.~\eqref{eq:max.frac}. Importantly, the mechanism enabling long-distance transport here differs from that in Result 2. There, the decoherence-free subspace comprises an extensive set of states, allowing the dynamics to remain confined within it. In contrast, the present decoherence-free subspace consists solely of the single dark state $|\psi\rangle$, preventing the system from evolving fully within the subspace. Nevertheless, when the initial particle density is below $\bar{n}$, the gain process drives the system toward $|\psi\rangle$, making the transport effectively immune to losses and thereby enabling nearly perfect transport.

Similarly, we figure out the transport limit $d_l$ as
\begin{equation}\label{eq: distance-2}
    d_l=(NJ\varphi\zeta(\alpha-\alpha_{\varepsilon}-D+1)B(\tau_c))^{1/\alpha_{\varepsilon}}.
\end{equation}
In the dilute limit, the distance $d_l$ can be rewritten as
\begin{equation}
    d_l=\left(\frac{\gamma_2|\Lambda|J\varphi\zeta(\alpha-\alpha_{\varepsilon}-D+1)}{(\Delta\gamma)^2}\right)^{1/\alpha_{\varepsilon}}.
\end{equation}
Therefore, the transport limit $d_l$ can be comparable with the system size, allowing particles to reach any site in the lattice---even for arbitrarily small gain rates $\gamma_2$---provided the initial average density is below $\bar{n}$. To be specific, in a three-dimensional lattice, if we take $\gamma_2/\Delta\gamma\sim{O}(|\Lambda|^{-1/2})$ and $J/\Delta\gamma\sim{O}(1)$, then the transport limit satisfies $d_l\sim{O}(|\Lambda|^{1/2})$, which is larger than the system length ${O}(|\Lambda|^{1/3})$ in the thermodynamic limit. Combining the results above, we conclude that the particle gain plays a nontrivial role in the dilute limit, more than simply compensating for the particle loss.

In addition, our results also hold for local Lindbladian dynamics
 \begin{equation}\label{eq: Lindblad_2}
   \frac{d \rho_t}{d t} = i [\rho_t,H] +\!\! \sum\limits_{\substack{i\in\Lambda\\\alpha=1,2}} \frac{\gamma_{\alpha}}{2}
   (2 L_i^{\alpha} \rho_t L_i^{\alpha\dagger} - L_i^{\alpha\dagger} L_i^{\alpha} \rho_t - \rho_t
   L_i^{\alpha\dagger} L_i^{\alpha}),
\end{equation}
where $L_i^1:=b_i$ and $L_i^2:=b_i^{\dagger}$.
The Lindbladian can be realized by coupling the system to two baths with different chemical potentials: one higher and one lower than the system.

\subsection{Transport Probability}
\noindent \textbf{Result 4}. 
Assume that the initial boson number outside region $X$ is $Nx_{X^c}(0)$. In the presence of local $n$-body loss, the probability of finding $N(x_{X^c}(0)+\mu)$ bosons in a disjoint region $Y$ in a given time $\tau$ is upper bounded by
\begin{equation}\label{eq: closed_P}
   \!\!\!\! P_{x_Y(\tau) \geq x_{X^c}(0)+\mu}\!\leq\frac{J \varphi \zeta (\alpha - \alpha_{\varepsilon} - D + 1) \tau}{\mu d_{X Y}^{\alpha_{\varepsilon}}},
\end{equation}
where the ratios $x_{X^c}(0)$ and $\mu$ are of the order ${O}(1)$.

It is worth noting that Result 2 can be interpreted as a corollary of Result 4 by following the same procedure in Ref. \cite{Vu_2024}. To be specific, by choosing an arbitrary constant $\mu'\in(0,\mu)$ and applying Markov's inequality, one can show that
\begin{equation}\label{eq: lower_prob}
    P_{x_Y(\tau) \ge x_{X^c}(0)+\mu'}\geq\frac{x_Y(\tau)-x_{X^c}(0)-\mu'}{1-x_{X^c}(0)-\mu'}.
\end{equation}
When the particle transport satisfies $x_Y(\tau)\geq x_{X^c}(0)+\mu$, we obtain
\begin{equation}
    \tau\geq\frac{\mu'(\mu-\mu')}{\mu[1-x_{X^c}(0)-\mu']}\kappa_1^{\varepsilon}d_{XY}^{\min(1,\alpha-D-\varepsilon)},
\end{equation}
by combining Eqs.~\eqref{eq: closed_P} and \eqref{eq: lower_prob}. Since the factor $\mu'(\mu-\mu')/\{\mu[1-x_{X^c}(0)-\mu']\}\leq1$ can be regarded as an $O(1)$ constant, the upper bound of the probability in Eq.~\eqref{eq: closed_P} shows fundamentally the same limit in particle transport as Result 2 on macroscopic transport. In comparison with the result for closed quantum systems~\cite{Vu_2024}, we conclude that dissipation does not generally enhance transport speed.

\section{Discussion}
To summarize, by establishing the optimal transport theory in open quantum systems, we have constructed the macroscopic particle transport theory of dissipative bosonic systems with long-range hopping and long-range interactions. By defining the cost function and generalized Wasserstein distance in open quantum systems, we have rigorously obtained the lower bound for the transport time of the bosonic systems with local particle loss and gain. More interestingly, we generally demonstrate that the presence of decoherence-free subspaces facilitates long-distance transport processes. It is worth noting that the above analysis and findings for bosonic systems are similarly applicable to fermionic systems. 

A significant prospect is further exploring the optimal Lieb-Robinson bound for dissipative bosonic systems with long-range interaction and long-range hopping. However, due to the unbounded nature of bosons, the optimal bound may be difficult to figure out. Additionally, one can consider a more general nonlocal dissipation rate $\gamma_{ij}$ satisfying a power-law decay is meaningful. However, the analysis goes beyond the optimal transport theory. A tighter upper bound for the probability of microscopic transport in the presence of $n$-body loss in Result 4 is also expected as a future work.

To bridge theory and experiments, we have also analyzed the experimental feasibility for validating our findings. A platform based on Rydberg-dressed neutral atoms trapped in optical lattices or tweezer arrays~\cite{PhysRevA.89.011402,PhysRevLett.124.063601,PhysRevA.111.L011305} makes it possible to realize all parts of our model, including the required preparation and measurements. The details of the feasible quantum simulation procedure are shown in Sec. \ref{sec.E}.

\section{Methods}\label{sec:method}
Before going into detailed proofs, we briefly explain the optimal transport theory, which concerns the optimal way to transport distributions.
The main concept in the theory is the Wasserstein distance \cite{Villani.2008}, which is defined as the minimum transport cost over all feasible plans that redistribute a given distribution to a desired one. Specifically, the Wasserstein distance between two finite-dimensional distributions $\textbf{x}$ and $\textbf{y}$ is defined as
\begin{equation}\label{eq: tradition_WD-1}
W(\textbf{x},\textbf{y})\assign\min_{\pi}\sum_{m,n}\pi_{mn}c_{mn},
\end{equation}
where $\pi$ denotes a joint probability distribution of $\textbf{x}$ and $\textbf{y}$ satisfying $\sum_m\pi_{mn}=x_n$ and $\sum_n\pi_{mn}=y_m$.
The quantity $\pi_{ij}$ can be interpreted as the amount of probability transported from $x_j$ to $y_i$. The cost function $c_{mn}$ is positive and symmetric as $c_{mn}=c_{nm}$, which satisfies $c_{mn}+c_{np}\geq c_{mp}$ for any $m,n,p$. These properties guarantee that the Wasserstein distance also satisfies the triangle inequality.
Notably, the Kantorovich-Rubinstein duality~\cite{Villani.2008,Vu_2024} derives another variational form of the Wasserstein distance,
\begin{equation}\label{eq:KR.dual}
W(\textbf{x},\textbf{y})=\max_{\textbf{w}}\textbf{w}^\top(\textbf{y}-\textbf{x}),
\end{equation}
where the maximum is over all vectors $\textbf{w}$ satisfying $|w_m-w_n|\le c_{mn}$.
 
The Wasserstein distance defined in Eq.~\eqref{eq: tradition_WD-1} applies to balanced distributions $\textbf{x}$ and $\textbf{y}$. That is, $\|\textbf{x}\|_1$ and $\|\textbf{y}\|_1$ must be identical, where $\|\textbf{x}\|_1\assign\sum_nx_n$.
However, in the presence of particle loss, the particle number decays, and we may encounter cases where the $1$-norms of $\textbf{x}$ and $\textbf{y}$ differ. Hence, the conventional distance cannot be straightforwardly applied in general. To address this issue, we define a generalized Wasserstein distance between two vectors $\textbf{x}$ and $\textbf{y}$ satisfying $\|\textbf{x}\|_1\geq\|\textbf{y}\|_1$ as
\begin{equation}\label{eq: generalized_WD}
    \widetilde{W}(\textbf{x},\textbf{y})\assign \min_{\textbf{x}\succeq\textbf{x}'\succeq\textbf{0},\textbf{y}'\succeq\textbf{y},\|\textbf{x}'\|_1=\|\textbf{y}'\|_1} W(\textbf{x}',\textbf{y}'),
\end{equation}
where $\textbf{x}\succeq\textbf{x}'$ indicates that $x_n\geq x_n'$ for all $n$. We can further prove that the generalized Wasserstein distance satisfies the triangle inequality (the proof is provided in Section II of Supplemental Information).

\subsection{Outline of the proof for Result 1}\label{sec: one-loss}
Here we consider the case of local one-body loss (i.e., $L_i=b_i)$.
The dynamics can be rewritten as
\begin{equation}\label{eq: local_L}
  \frac{d \rho_t}{d t} = - i [H, \rho_t] + \frac{\gamma}{2}\sum_{i} 
  (2 b_i \rho_t b_i^{\dagger} - b_i^{\dagger} b_i \rho_t - \rho_t
  b_i^{\dagger} b_i).
\end{equation}
By defining the particle ratio in site $i$ at time $t$ as $x_i(t) \assign \tmop{tr}(b_i^\dagger b_i\rho_t) / N$, the time evolution of $x_i(t)$ can be derived from Eq.~\eqref{eq: local_L} as
\begin{equation}
    \dot{x}_i(t)=\frac{1}{N} \sum_{j, j \neq i} 2 J_{i j} (t) \tmop{Im} [\tmop{tr}
  (b_j^{\dagger} b_i \rho_t)] - \gamma x_i(t).
\end{equation}
By solving this equation, we obtain
\begin{equation}\label{eq: dynamics-one}
    x_i (\tau) = x_i (0) e^{- \gamma \tau} + \int_0^{\tau}dt  \sum_{j,
   j \neq i} \tilde{\phi}_{i j} (t),
\end{equation}
with the definition
\begin{equation}
    \tilde{\phi}_{i j} (t) \assign 2 J_{i j} (t)e^{\gamma (t - \tau)} \tmop{Im} [\tmop{tr} (b_j^{\dagger} b_i \rho_t)] / N.
\end{equation}
Here we can see that the particle current originates from the coherence in the density matrix, which is intrinsically different from the transport mechanism in the classical rate equation, such as the reaction-diffusion equation. From Eq.~\eqref{eq: dynamics-one}, we can see the dynamics of the particle number at site $i$ can be separated into two parts. The first term on the right-hand side of Eq.~\eqref{eq: dynamics-one} represents the particle decay due to the one-body loss since the total number also exponentially decays with time:
\begin{equation}\label{eq:tot_N}
    \sum_i x_i(t)=\sum_i x_i(0)e^{-\gamma t} =e^{-\gamma t}.
\end{equation}
The second term on the right-hand side of Eq.~\eqref{eq: dynamics-one} is the current flow from all the sites other than $i$. Therefore, $ \tilde{\phi}_{i j}$ represents the current from the site $j$ to the site $i$ at time $t$.

Since we consider the case where a fraction $\mu$ of the total bosons must be transported from $X$ to $Y$ within the time period $\tau$, the following relation should hold:
\begin{align}
    x_{Y}(\tau)-y_{X^c}(\tau)\geq \mu,
\end{align}
where $X^c\assign\Lambda\setminus X$ and $y_{X^c}(\tau)=e^{-\gamma\tau}x_{X^c}(0)$ denotes the density at time $\tau$ in region $X^c$ in the presence of one-body loss by switching off the Hamiltonian. We consider the Wasserstein distance between the initial distribution $e^{-\gamma\tau}\textbf{x}_0$ and the final distribution $\textbf{x}_\tau$, with the cost function is defined as $c_{ij}=\|i-j\|^{\alpha_{\varepsilon}}$.
Any feasible coupling $\pi$ should satisfy $\sum_i\pi_{ij}=e^{-\gamma\tau}x_j(0)$ and $\sum_j\pi_{ij}=x_i(\tau)$. Accordingly, we obtain
\begin{align}
    \mu&\le x_Y(\tau)-e^{-\gamma\tau}x_{X^c}(0)\notag\\
    &=\sum_{i\in Y,j\in\Lambda}\pi_{ij}-\sum_{i\in\Lambda,j\in X^c}\pi_{ij}\notag\\
    &=\sum_{i\in Y,j\in X}\pi_{ij}-\sum_{i\in Y^c,j\in X^c}\pi_{ij}
    \leq\sum_{i\in Y,j\in X}\pi_{ij}.
\end{align}
Therefore, the Wasserstein distance can be lower bounded by
\begin{align}
    W(e^{-\gamma\tau}\textbf{x}_0,\textbf{x}_\tau)\ge\min_{i\in Y,j\in X}c_{ij}\sum_{i\in Y,j\in X}\pi_{ij}\ge\mu d_{XY}^{\alpha_{\varepsilon}}.\label{eq:Wdist.lb}
\end{align}
On the other hand, applying the Kantorovich-Rubinstein duality yields
\begin{align}
	W(e^{-\gamma\tau}\textbf{x}_0,\textbf{x}_\tau)&=\max_{\textbf{w}}\sum_iw_i\int_0^{\tau}dt  \sum_{j(\neq i)}\tilde{\phi}_{i j} (t)\notag\\
	&=\frac{1}{2}\max_{\textbf{w}}\sum_{i\neq j}\int_0^{\tau}dt\,\tilde{\phi}_{i j} (t)(w_i-w_j)\notag\\
	&\le \frac{1}{2}\sum_{i\neq j}\int_0^{\tau}dt\,c_{ij}|\tilde{\phi}_{i j} (t)|,\label{eq:Wdist.ub}
\end{align}
where we use the relations $\tilde{\phi}_{ij}(t)=-\tilde{\phi}_{ji}(t)$ and $|w_i-w_j|\le c_{ij}$. Meanwhile, the current $\tilde{\phi}_{ij}$ can be upper bounded by
\begin{align}
    | \tilde{\phi}_{i j} (t) | & = \frac{J_{i j} (t)}{N} | \tmop{tr}
  [(b_j^{\dagger} b_i - b_i^{\dagger} b_j) \rho_t] | e^{\gamma (t - \tau)}\nonumber\\
  & \leq J_{i j} (t) [x_i (t) + x_j (t)] e^{\gamma (t - \tau)} .
\end{align}
Consequently, the Wasserstein distance can be upper bounded as
\begin{align}
  W(e^{-\gamma\tau}\textbf{x}_0,\textbf{x}_{\tau}) &
   \leq \sum_{i \neq j} J_{i j} c_{i j} \int_0^{\tau}d t\, x_i (t)
  e^{\gamma (t - \tau)}  \nonumber\\
  &\leq\sum_i\int_0^{\tau}dt \sum_{l=1}^{\infty}\frac{J\varphi x_i(t)e^{\gamma(t-\tau)}}{l^{\alpha-\alpha_{\varepsilon}-D+1}}\nonumber\\
  & = \tau e^{- \gamma \tau} J \varphi \zeta (\alpha - \alpha_{\varepsilon}
  - D + 1),\label{eq:cost.ub}
\end{align}
where we use Eq.~\eqref{eq:tot_N} to obtain the last inequality. Therefore, by combining Eq.~\eqref{eq:cost.ub} with Eq.~\eqref{eq:Wdist.lb}, we obtain the lower-bound condition for the transport time $\tau$ as
\begin{equation}
    \tau e^{- \gamma \tau} \geq\kappa_1^{\varepsilon} d_{X
  Y}^{\alpha_{\varepsilon}}. 
\end{equation}
This completes the proof.

\subsection{Outline of the proof for Result 2}
For the case of local multi-body loss (i.e., $L_i=b_i^n$ and $n>1$), an analogy of the analysis in the previous section fails since we cannot exactly solve the dynamics of $x_i(t)$ anymore. Nevertheless, with the help of the generalized Wasserstein distance \eqref{eq: generalized_WD}, we can resolve the problem.
In the presence of particle loss, the dynamics of $x_i(t)$ can be written in a general form of
\begin{equation}
    \dot{x}_i(t)=-d_i(t)+\sum_{j(\neq i)}\phi_{ij}(t).
\end{equation}
Here, $d_i(t)\assign-(\gamma/2N)\tmop{tr}(\hat{d}_i\rho_t)$ represents the local loss rate of bosons at site $i$, $\phi_{ij}(t)\assign 2N^{-1}J_{i j} (t) \tmop{Im} [\tmop{tr}(b_j^{\dagger} b_i \rho_t)]$ is the particle current exchanged between sites due to hopping, and $\hat{d}_i\assign 2(b_i^\dagger)^n\hat{n}_ib_i^n-\hat{n}_i(b_i^\dagger)^nb_i^n-(b_i^\dagger)^nb_i^n\hat{n}_i$. Crucially, we can show that the loss rate $d_i(t)$ is nonnegative for any $i$ as follows:
\begin{equation}
    d_i(t)=\frac{\gamma}{N}\sum_{\vec{N}}n\Theta(n_i-n)\tmop{tr}(\Pi_{\vec{N}}\rho_t)\prod_{k=0}^{n-1}(n_i-k)\geq0,
\end{equation}
where $\Pi_{\vec{N}}:=|\vec{N}\rangle\langle\vec{N}|$ is the projection operator onto the eigenstates of number operators, i.e., $\hat{n}_i|\vec{N}\rangle=n_i|\vec{N}\rangle$, and $\Theta(n_i-n)$ is the Heaviside function. We consider the generalized Wasserstein distance between $\textbf{x}_0$ and $\textbf{x}_{\tau}$ with the cost function $c_{ij}=\|i-j\|^{\alpha_\varepsilon}$.
Define $\tilde{x}_i\assign x_i(\tau)+\int_0^\tau dt\,d_i(t)=x_i(0)+\int_0^\tau dt\,\sum_{j(\neq i)}\phi_{ij}(t)$.
Note that $\tilde{\textbf{x}}\succeq\textbf{x}_\tau$ and $\|\textbf{x}_0\|_1=\|\tilde{\textbf{x}}\|_1$ since the current $\phi_{ij}(t)$ is antisymmetric. Applying the Kantorovich-Rubinstein duality, the distance $\widetilde{W}(\textbf{x}_0,\textbf{x}_\tau)$ can be upper bounded as follows:
\begin{align}
    \widetilde{W}(\textbf{x}_0,\textbf{x}_{\tau})&=\min_{\textbf{x}_0\succeq\textbf{x}'\succeq\textbf{0},\textbf{x}''\succeq\textbf{x}_\tau,\|\textbf{x}'\|_1=\|\textbf{x}''\|_1}\max_{\textbf{w}}\textbf{w}^\top(\textbf{x}''-\textbf{x}')\notag\\
    &\le \max_{\textbf{w}}\textbf{w}^\top(\tilde{\textbf{x}}-\textbf{x}_0)\notag\\
    &\le \frac{1}{2}\int_0^\tau dt\,\sum_{i\neq j}c_{ij}|\phi_{ij}(t)|,\label{eq:gWdist.ub}
\end{align}
where the first equality is obtained by substituting the Kantorovich-Rubinstein duality \eqref{eq:KR.dual} into the generalized Wasserstein distance \eqref{eq: generalized_WD} and the second line is achieved by assigning $\textbf{x}'=\textbf{x}_0$ and $\textbf{x}''=\tilde{\textbf{x}}$. Similarly, its lower bound can be established as
\begin{align}
    \widetilde{W}(\textbf{x}_0,\textbf{x}_{\tau})&=\min_{\textbf{x}_0\succeq\textbf{x}'\succeq\textbf{0}, \textbf{x}''\succeq\textbf{x}_{\tau},\|\textbf{x}'\|_1=\|\textbf{x}''\|_1}\min_{\pi}\sum_{i,j}\pi_{ij}c_{ij}\nonumber\\
    &\geq\mu d_{XY}^{\alpha_{\varepsilon}},\label{eq:gWdist.lb}
\end{align}
where the last equality is obtained using the following fact:
\begin{align}
    \mu&\leq x_{Y}(\tau)-x_{X^c}(0)
    \le x_{Y}''-x_{X^c}'\notag\\
    &=\sum_{i\in Y,j\in X}\pi_{ij}-\sum_{i\in Y^c,j\in X^c}\pi_{ij}\leq\sum_{i\in Y,j\in X}\pi_{ij}.
\end{align}
Following the same procedure as in Eq.~\eqref{eq:cost.ub} and noting that $\|\textbf{x}_t\|_1\le 1$, we can show that
\begin{equation}
	\frac{1}{2}\int_0^\tau dt\,\sum_{i\neq j}c_{ij}|\phi_{ij}(t)|\le \tau J\varphi\zeta(\alpha-\alpha_\varepsilon-D+1).\label{eq:gcost.ub}
\end{equation}
Combining Eqs.~\eqref{eq:gWdist.ub}, \eqref{eq:gWdist.lb}, and \eqref{eq:gcost.ub} yields the desired result. 

Based on the above analysis, we conclude that the bounds for multi-body loss remain the same as those in closed quantum systems.

Finally, we note that this generalized Wasserstein distance can also be applied to the case with one-body loss. However, the bound in Eq.~\eqref{eq:gcost.ub} is not tight since the equality requires that no particles are lost. This condition can be satisfied in Result 2, where a decoherence-free subspace emerges from the multi-body loss. However, there does not exist a nontrivial decoherence-free subspace in Result 1, so we introduce a new Wasserstein distance $W(e^{-\gamma\tau}\textbf{x}_0,\textbf{x}_{\tau})$ to obtain a tighter bound.
\subsection{Outline of the proof for Result 3}
Under the Lindbladian dynamics~\eqref{eq: Lindblad_2}, the dynamics of $x_i$ is given by
\begin{align}\label{eq: gal}
    \!\!\!\!\dot{x}_i(t)=\frac{1}{N}\sum_{j, j \neq i}& 2 J_{i j} (t) \tmop{Im} [\tmop{tr}
  (b_j^{\dagger} b_i \rho_t)] \nonumber\\
  &- \gamma_1 x_i(t)+\gamma_2 \left[x_i(t)+\frac{1}{N}\right].
\end{align}
Here, $\gamma_{1}$ and $\gamma_{2}$ represent the loss and gain rates, respectively. By defining
\begin{equation}
\tilde{x}_i(t):=x_ie^{(\gamma_1-\gamma_2)t}-\frac{\gamma_2}{N(\gamma_1-\gamma_2)} e^{(\gamma_1-\gamma_2)t},    
\end{equation}
we obtain the evolution of $\tilde{x}_i$ as
\begin{equation}
    \frac{d}{dt}\tilde{x}_i(t)=\frac{1}{N}\sum_{j, j \neq i} 2 J_{i j} (t) \tmop{Im} [\tmop{tr}
  (b_j^{\dagger} b_i \rho_t)]e^{(\gamma_1-\gamma_2)t}.
\end{equation}
Therefore, the solution to $x_i$ takes the form of
\begin{align}
    x_i(\tau)&=x_i(0)e^{-\Delta\gamma \tau}+\frac{\gamma_2}{N\Delta\gamma}(1-e^{-\Delta\gamma\tau})\\
    &+\int_0^{\tau}dt\frac{1}{N}\sum_{j, j \neq i} 2 J_{i j} (t) \tmop{Im} [\tmop{tr}
  (b_j^{\dagger} b_i \rho_t)]e^{\Delta\gamma(t-\tau)},\nonumber
\end{align}
where $\Delta\gamma:=\gamma_1-\gamma_2$. The first and second terms on the right-hand side represent the joint effect of the particle gain and loss, while the third term represents the particle transport. The total number evolves with time as
\begin{align}\label{eq: sum_2}
    \sum_ix_i(\tau)&=\sum_ix_i(0)e^{-\Delta\gamma\tau}+\frac{\gamma_2}{\Delta\gamma}\mathcal{N}^{-1}(1-e^{-\Delta\gamma\tau})\nonumber\\
    &=e^{-\Delta\gamma\tau}+\frac{\gamma_2}{\Delta\gamma}\mathcal{N}^{-1}(1-e^{-\Delta\gamma\tau}).
\end{align}
Therefore, we can define an effective particle current for transport
\begin{equation}
    \tilde{\phi}_{ij}(t):=2 J_{i j} (t) \tmop{Im} [\tmop{tr}(b_j^{\dagger} b_i \rho_t)]e^{\Delta\gamma(t-\tau)}/N.
\end{equation}
Similar to the analysis in Sec.~\ref{sec: one-loss}, we here consider the Wasserstein distance between the distribution $\textbf{y}_{\tau}$ and the final distribution $\textbf{x}_{\tau}$ with the cost function defined as $c_{ij}=\|i-j\|^{\alpha_{\varepsilon}}$. Here the distribution $\textbf{y}_{\tau}$ satisfies
\begin{equation}
    y_i(\tau)=y_i(0)e^{-\Delta\gamma\tau}+\frac{\gamma_2}{N\Delta\gamma}(1-e^{-\Delta\gamma\tau}).
\end{equation}
 We note that $\textbf{y}$ represents the evolution of the particle distribution in the absence of hopping. If a fraction $\mu$ of the total particles is transported from $X$ to $Y$ in a time period $\tau$, we have the following relation:
\begin{align}
    x_{Y}(\tau)-y_{X^c}(\tau)\geq \mu.
\end{align}
Then a feasible coupling $\pi_{ij}$ should satisfy $\sum_i\pi_{ij}=y_j(\tau)$ and $\sum_j\pi_{ij}=x_i(\tau)$. Accordingly, we have
\begin{align}
    \mu&\le x_Y(\tau)-y_{X^c}(\tau)
    \leq\sum_{i\in Y,j\in X}\pi_{ij}.
\end{align}
Therefore, the Wasserstein distance can be similarly lower bounded by
\begin{equation}\label{eq: lower_gl}
    W(\textbf{y}_{\tau},\textbf{x}_{\tau})\geq\min_{i\in Y,j\in X}c_{ij}\sum_{i\in Y,j\in X}\pi_{ij}\ge\mu d_{XY}^{\alpha_{\varepsilon}}.
\end{equation}
To obtain the upper bound, we still apply the Kantorovich-Rubinstein duality and obtain
\begin{align}
    W(\textbf{y}_{\tau},\textbf{x}_{\tau})\leq\frac{1}{2}\sum_{i\neq j}\int_0^{\tau}dtc_{ij}|\tilde{\phi}_{ij}(t)|,
\end{align}
where the analysis is the same as the one in Eq.~\eqref{eq:Wdist.ub}. The current can also be upper bounded by
\begin{align}
    | \tilde{\phi}_{i j} (t) | & = \frac{J_{i j} (t)}{N} | \tmop{tr}
  [(b_j^{\dagger} b_i - b_i^{\dagger} b_j) \rho_t] | e^{\Delta\gamma (t - \tau)}\nonumber\\
  & \leq J_{i j} (t) (x_i (t) + x_j (t)) e^{\Delta\gamma (t - \tau)} .
\end{align}
Hence, we obtain an upper bound for the summation of the cost function, given by
\begin{align}\label{eq: summation of cost function}
    W(\textbf{y}_{\tau},\textbf{x}_{\tau})& \leq\sum_i\int_0^{\tau}dt \sum_{l=1}^{\infty}\frac{J\varphi x_i(t)e^{\Delta\gamma(t-\tau)}}{l^{\alpha-\alpha_{\varepsilon}-D+1}}\nonumber\\
  & = A(\tau) J \varphi \zeta (\alpha - \alpha_{\varepsilon}- D + 1),
\end{align}
where
\begin{equation}
    A(\tau):= \tau e^{-\Delta\gamma\tau}+\frac{\gamma_2}{\Delta\gamma}\mathcal{N}^{-1}\left(\frac{1-e^{-\Delta\gamma\tau}}{\Delta\gamma}-\tau e^{-\Delta\gamma\tau}\right).
\end{equation}
Here we use Eq.~\eqref{eq: sum_2} in the second inequality of Eq.~\eqref{eq: summation of cost function}. By combining Eq.~\eqref{eq: summation of cost function} with Eq.~\eqref{eq: lower_gl}, we obtain the final result
\begin{equation}
    \tau e^{-\Delta\gamma\tau}+\frac{\gamma_2}{\Delta\gamma}\mathcal{N}^{-1}\left(\frac{1-e^{-\Delta\gamma\tau}}{\Delta\gamma}-\tau e^{-\Delta\gamma\tau}\right)\geq\kappa_1^{\varepsilon}d_{XY}^{\alpha_{\varepsilon}},
\end{equation}
which completes the proof.

\subsection{Outline of the proof for Result 4}
Under the Lindbladian dynamics \eqref{eq: Lindblad}, the evolution equation of the probability distribution $p_{\vec{N}}(t):=\langle \vec{N} |
  \rho_t | \vec{N} \rangle$ is generally shown as
\begin{align}\label{eq: prob-open}
    \dot{p}_{\vec{N}} (t)&=i \sum_{i \neq j} J_{i j} (t) \sqrt{n_i n_j'} (\langle \vec{N} |
  \rho_t | \vec{N}' \rangle - \langle \vec{N}' | \rho_t | \vec{N} \rangle)\nonumber\\
  &+\gamma \sum_i \frac{(n_i+n)!}{n_i!} p_{\vec{N}^{+}_i} - \gamma\sum_i\frac{n_i!}{(n_i-n)!} p_{\vec{N}},
\end{align}
where $\vec{N}^{+}_i$ is the state with the particle distribution $\{n_i^{+}\}_{i\in\Lambda}$ satisfying ${n^{+}_i} = {n_i} + n$ and ${n^{+}_k} = {n_k}\left( {k \ne i} \right)$ compared with the distribution $\{n_i\}_{i\in\Lambda}$ of the state $\vec{N}$, i.e., $\hat{n}_i|\vec{N}\rangle=n_i|\vec{N}\rangle$. The particle distribution $\{n_i'\}_{i\in\Lambda}$ of the state $\vec{N}'$ satisfies $n_i'=n_i-1$ and $n_j'=n_j+1$.
For a general case with arbitrary $n$-body loss, the traditional Wasserstein distance breaks down since the quantum jump process changes the support of the probability distribution. However, since the particle loss is local, we only consider two types of neighboring states. The first type is for any two vectors $\vec{N}$ and $\vec{M}$ that satisfy $n_i-m_i=-(n_j-m_j)=\pm 1$ for some $i,j$, and $n_k=m_k$ for all $k\neq i,j$. For these states, we define $c_{\vec{M}\vec{N}}=c_{\vec{N}\vec{M}}=\|i-j\|^{\alpha_{\varepsilon}}$. The second type is for any two vectors $\vec{N}$ and $\vec{M}$ that satisfy $n_i-m_i=n$ for some $i$, and $n_k=m_k$ for all $k\neq i$. For these states, we define the cost as $c_{\vec{M}\vec{N}}=0$. Then, for general vectors $\vec{N}$ and $\vec{M}$ that satisfy $\|\vec{M}\|_1\leq\|\vec{N}\|_1$, we can define the cost between them as the shortest path cost over all possible paths connecting the two states as follows:
\begin{equation}
    c_{\vec{M}\vec{N}}=\min\sum_{k=1}^K c_{\vec{N}_k\vec{N}_{k-1}},
\end{equation}
where $\vec{N}_0=\vec{N}$, $\vec{N}_K=\vec{M}$, and $\vec{N}_k$ and $\vec{N}_{k-1}$ are neighboring states for all $1\leq k\leq K$. For general states $\vec{M}$ and $\vec{N}$ satisfying $\|\vec{M}\|_1>\|\vec{N}\|_1$, we define the cost $c_{\vec{M}\vec{N}}=\Omega+1$ where $\Omega:=\max_{\|\vec{M}\|_1\leq\|\vec{N}\|_1}c_{\vec{M}\vec{N}}$. One can show that the cost we defined satisfies the triangular inequality. The time evolution can be expressed as
\begin{equation}
    \dot{p}_{\vec{N}}(t)=\sum_{i\neq j}\phi_{\vec{N}\vec{N}'}(t)+\sum_i[\varphi_{\vec{N}\vec{N}^{+}_i}(t)+\varphi_{\vec{N}\vec{N}^{-}_i}(t)],
\end{equation}
where
\begin{equation}
\phi_{\vec{N}\vec{N}'}:=iJ_{i j} \sqrt{n_i n_j'} (\langle \vec{N} |\rho_t | \vec{N}' \rangle - \langle \vec{N}' | \rho_t | \vec{N} \rangle),
\end{equation}
$\varphi_{\vec{N}\vec{N}^{+}_i}:=\gamma\frac{(n_i+n)!}{n_i!}p_{\vec{N}^{+}_i}$, and $\varphi_{\vec{N}\vec{N}^{-}_i}:=-\gamma \frac{n_i!}{(n_i-n)!}p_{\vec{N}}$. The state $\vec{N}_i^{-}$ is obtained by setting $n_i^{-}=n_i-n$ and $n_j^{-}=n_j$ for all $j\neq i$. We note that $\varphi_{\vec{N}\vec{N}^{+}_i}$ and  $\varphi_{\vec{N}\vec{N}^{-}_i}$ satisfy $\varphi_{\vec{N}\vec{N}^{+}_i}=-\varphi_{\vec{N}^{+}_i\vec{N}}$ and $\varphi_{\vec{N}\vec{N}^{-}_i}=-\varphi_{\vec{N}^{-}_i\vec{N}}$, respectively. By applying the following inequality:
\begin{equation}
    W(\textbf{p},\textbf{q})\leq\max_{f_i-f_j\leq c_{ij}}\textbf{f}^\top(\textbf{p}-\textbf{q}),
\end{equation}
which holds even for asymmetric $c_{ij}$, we bound the Wasserstein distance $W(\textbf{p}_0,\textbf{p}_{\tau})$ as below:
\begin{align}
    W(\textbf{p}_0,\textbf{p}_{\tau})&\leq \max_{f_{\vec{M}}-f_{\vec{N}}\leq c_{\vec{M}\vec{N}}}\textbf{f}^\top(\textbf{p}_0-\textbf{p}_{\tau})\nonumber\\
    &\leq\max_{f_{\vec{M}}-f_{\vec{N}}\leq c_{\vec{M}\vec{N}}}\int_0^{\tau}dt\Bigg[\frac{1}{2}\sum_{\vec{N},\vec{N}'}(f_{\vec{N}}-f_{\vec{N}'})\phi_{\vec{N}\vec{N}'}\nonumber\\
    &+\sum_{\vec{N}}\sum_i(f_{\vec{N}}-f_{\vec{N}^{+}_i})\varphi_{\vec{N}\vec{N}^{+}_i}\Bigg],
\end{align}
where $\textbf{p}_0=[p_{\vec{N}}(0)]$ and $\textbf{p}_{\tau}=[p_{\vec{N}}(\tau)]$ denote the probability distributions over particle number configurations $\vec{N}$ across all lattice sites at the initial and final times, respectively. Since $|f_{\vec{N}}-f_{\vec{N}'}|\leq c_{\vec{N}\vec{N}'}$ and $f_{\vec{N}}-f_{\vec{N}^{+}_i}\leq0$, we subsequently obtain
\begin{equation}
     W(\textbf{p}_0,\textbf{p}_{\tau})\leq\frac{1}{2}\int_0^{\tau}dt\sum_{\vec{N},\vec{N}'}c_{\vec{N}\vec{N}'}|\phi_{\vec{N}\vec{N}'}(t)|.
\end{equation}
Since the current $|\phi_{\vec{N}\vec{N}'}(t)|$ is the same as the one in closed quantum systems and has no relation to the dissipation, we can similarly bound $W(\textbf{p}_0,\textbf{p}_{\tau})$ in the same way as in closed systems given by (see Supplemental Section I for derivations)
\begin{equation}\label{eq: distances}
    W(\textbf{p}_0,\textbf{p}_{\tau})\leq N \tau J \varphi \zeta (\alpha - \alpha_{\varepsilon} - D + 1).
\end{equation}
On the other hand, the Wasserstein distance can be rewritten as
\begin{equation}
     W (\textbf{p}_0, \textbf{p}_{\tau})  =  \min_{\pi} \sum_{\vec{M},
  \vec{N}} c_{\vec{M} \vec{N}} \pi_{\vec{M} \vec{N}},
\end{equation}
where the joint probability distribution $\pi_{\vec{M} \vec{N}}$ satisfies $\sum_{\vec{N}}\pi_{\vec{M} \vec{N}}=p_{\vec{M}}(\tau)$ and $\sum_{\vec{M}}\pi_{\vec{M} \vec{N}}=p_{\vec{N}}(0)$. Since the cost is nonzero if and only if the transport process, rather than particle loss, takes place, the analysis is identical to that for closed quantum systems, and we can obtain (see Supplemental Section I for the derivation)
\begin{equation}
  W (\textbf{p}_0, \textbf{p}_{\tau}) \geqslant
  \mu N d_{X Y}^{\alpha_{\varepsilon}} P_{x_Y(\tau) \geqslant x_{X^c}(0) +\mu}.\label{eq:lower-WD}
\end{equation}
Combining Eqs.~\eqref{eq: distances} and \eqref{eq:lower-WD}, we achieve the same upper bound on the probability distribution as in closed quantum systems:
\begin{equation}
  P_{x_Y(\tau) \geqslant x_{X^c}(0) + \mu} \leq
  \frac{J \varphi \zeta (\alpha - \alpha_{\varepsilon} - D + 1) \tau}{\mu d_{X Y}^{\alpha_{\varepsilon}}},
\end{equation}
which completes the proof.

\subsection{Possible experimental situation}\label{sec.E}
Here we introduce the details of a possible experiment to measure our predicted results with quantum simulation based on the platform of Rydberg-dressed neutral atom arrays~\cite{PhysRevA.89.011402,PhysRevLett.124.063601,PhysRevA.111.L011305}.

\textit{Step 1. Preparation:}
For the general verification of our results, the system under study can be initialized by loading Rydberg atoms in an optical lattice or a tweezer array with near-unity filling, where the bosonic number distribution is encoded into the spin-1/2 system serving as a hard-core bosonic system. The optical lattice can be created by interfering laser beams which confines neutral atoms at specific sites~\cite{bloch2005ultracold}. Afterwards, acousto-optic deflectors and spatial light modulators can be used to shape the optical potential and precisely tune the particle distribution~\cite{kaufman2021quantum}, ensuring consistency with our theoretical setup.

\textit{Step 2. Manipulating particle transport in optical lattices:} The simulation of particle transport dynamics relies on realizing controllable long-range interactions, long-range hopping, local particle gain and loss in optical lattice experiments~\cite{RevModPhys.95.035002}, the core of which lies in the tuning of the laser fields. The dressing process involves applying an off-resonant laser field that hybridizes the atomic ground state with a highly excited Rydberg state, thereby creating Rydberg-dressed atoms with tunable long-range interactions set by the laser detuning and Rabi frequency~\cite{PhysRevA.89.011402,PhysRevLett.128.113602}. In addition, controllable long-range hopping amplitudes between lattice sites can be effectively achieved and tuned by Raman or microwave coupling~\cite{PhysRevLett.111.185302,PRXQuantum.3.020303}. Together, the tunable long-range interactions and engineered long-range hopping enable controlled particle transport across the lattice~\cite{wuster2011excitation,PhysRevResearch.6.033282}.  Additionally, controllable local particle loss can be realized by applying an additional laser that resonantly couples the Rydberg state to a short-lived intermediate level, which spontaneously decays to uncoupled ground or untrapped magnetic sublevels, thereby inducing an effective and tunable dissipation channel~\cite{w3x9-ll79}. Moreover, engineered local particle gain can be effectively achieved through the detuning of local Rydberg atoms which can be realized by local light AC Stark shifts~\cite{Oliveira2025,PhysRevA.106.052810,RevModPhys.82.2313}. 

\textit{Step 3. Measurement:} The particle transport is characterized through site-resolved detection of Rydberg atom state on each site. Before and after the transport process, single-shot fluorescence imaging provides the atom state distribution of the Rydberg atoms across the lattice~\cite{Labuhn2016,Scholl2023}, which encodes the particle number distribution over the lattice. By analyzing the detected signals over the predefined source region $X$ and target region $Y$, the number of particles transferred between them can be directly obtained. Finally, by repeating steps (1-3) multiple times and averaging the measured results, the error in the transport process can be reduced to a satisfactory range. 

Combining the methods above, Rydberg-dressed neutral atoms trapped in optical lattices provide the controllability required to support our theoretical findings~\cite{PhysRevA.82.033412,PhysRevLett.105.160404,PRXQuantum.4.020301,zeiher2016many}. In addition to this representative platform, there are also other promising candidates including cavity-QED systems with Bose-Einstein condensates~\cite{PhysRevX.8.011002}, exciton-polariton condensates in semiconductor microcavities~\cite{RevModPhys.82.1489}, and microwave-engineered arrays of circular Rydberg-atom ensembles for quantum simulation~\cite{PhysRevX.8.011032}.

\section*{Data availability}
Data sharing does not apply to this paper, as no datasets were generated or analyzed
during the current study.

\def\bibsection{\section*{References}} 


\begin{thebibliography}{101}
\bibitem [{dep()}]{dephasing_operator}%
  \BibitemOpen
  \href@noop {} {\bibinfo {title} {We note that the dephasing operator \( {L}_i = b_i^\dagger b_i \) commutes with the number operator and hence does not influence the particle transport.}}\BibitemShut {Stop}%
\end{thebibliography}

%

\section*{Acknowledgments}
{~}\\
We thank the referees for their constructive comments that helped improve our work.
{H.L.} is supported by Forefront Physics and Mathematics Program to Drive Transformation (FoPM), a World-leading Innovative Graduate Study (WINGS) Program, the University of Tokyo. {H.L.} also acknowledges JSPS KAKENHI Grant No.~JP24KJ0824. {C.S.} and {T.K.} acknowledge the Hakubi projects of RIKEN. {T.K.} was supported by JST PRESTO (Grant No.~JPMJPR2116), ERATO (Grant No.~JPMJER2302), and JSPS KAKENHI (Grants No.~JP23H01099 and No.~JP24H00071). {T.V.V.} was supported by JSPS KAKENHI (Grant No.~JP23K13032).

\section*{Author contributions}

{~}\\
H.L., C.S., T.K., and T.V.V. contributed to the conception of the work, the analysis and interpretation, and the preparation and revision of the manuscript.

\section*{Competing Interests}
{~}\\
The authors declare no competing interests.

\section*{Additional information}
{~}\\
\textbf{Supplementary information}

{~}\\

\clearpage{}
\onecolumngrid
\appendix
\begin{center}
	\large{Supplementary Information for}\\
	\textbf{``Macroscopic particle transport in dissipative long-range bosonic systems''}
\end{center}

\section{Macroscopic particle transport theory in closed quantum systems}\label{appendix-A}
In this section, we review the macroscopic particle transport theory in long-range closed quantum systems. Firstly, we show the proof of $\tau\geq\kappa_1^{\varepsilon}d_{XY}^{\alpha_{\varepsilon}}$. Since the density matrix follows the von Neumann equation, the dynamics of the boson number density $x_i(t)\coloneqq\text{tr}(n_i\rho_t)/N$ is given by
\begin{align}
    \dot{x}_i = \frac{1}{N} \sum_{j, j \neq i} 2 J_{i j} (t) \mathrm{Im}  [\mathrm{tr}(b_j^{\dagger} b_i \rho_t)]
    \eqqcolon\sum_{j(\neq i)}\phi_{ij}(t),
\end{align}
where $\phi_{ij}(t)$ is the current flowing from the site $j$ to the site $i$ satisfying $\phi_{ij}=-\phi_{ji}$. On the one hand, since a fraction $\mu$ of bosons should be transported from $X$ to $Y$ in the time period $\tau$, we have the relation
\begin{align}
    x_{Y}(\tau)-x_{X^c}(0)\geq \mu.
\end{align}
Here, $X^c\coloneqq\Lambda\setminus X$ is the complement of $X$. Since we consider the Wasserstein distance between the final distribution $\textbf{x}(\tau)$ and the initial distribution $\textbf{x}(0)$, the coupling $\pi_{mn}$ should satisfy $\sum_m\pi_{mn}=x_n(0)$ and $\sum_n\pi_{mn}=x_m(\tau)$. Accordingly, we obtain
\begin{align}\label{eq: lower-A3}
    x_Y(\tau)-x_{X^c}(0) \ =\sum_{i\in Y,j\in\Lambda}\pi_{ij} \ - \!\! \sum_{i\in\Lambda,j\in X^c}\pi_{ij} \ = \sum_{i\in Y,j\in X}\pi_{ij} \ -\!\! \sum_{i\in Y^c,j\in X^c}\pi_{ij}
    \ \leq\sum_{i\in Y,j\in X}\pi_{ij}. 
\end{align}
Hence, the Wasserstein distance can be lower bounded by
\begin{equation}\label{eq: lower-A4}
W(\textbf{x}_0,\textbf{x}_{\tau})\geq \min_{i\in Y,j\in X}c_{ij}\sum_{i\in Y,j\in X}\pi_{ij}\geq\mu d_{XY}^{\alpha_{\varepsilon}}.
\end{equation}
On the other hand, since the current can be upper bounded by
\begin{equation}
|\phi_{ij}|\leq|J_{ij}(t)|(x_i+x_j).
\end{equation}
With the Kantorovich-Rubinstein duality, we obtain
\begin{align}
W(\textbf{x}_0,\textbf{x}_{\tau})&\leq\max_{\|h\|_L\leq1}h^T(\textbf{x}(\tau)-\textbf{x}(0)) =\max_{\|h\|_L\leq1}\sum_i h_i\int_0^{\tau}dt\sum_{j(\neq i)}\phi_{ij}(t)\nonumber\\
     &\leq\frac{1}{2}\max_{\|h\|_L\leq1}\int_0^{\tau}dt\sum_{j\neq i}|\phi_{ij}||h_i-h_j|\leq\frac{1}{2}\int_0^{\tau}dt\sum_{j\neq i}c_{ij}|\phi_{ij}(t)|.
\end{align}
Here, we use $\|h\|_L\leq1$ to represent $|h_i-h_j|\leq c_{ij}$. Therefore, the upper bound of the Wasserstein distance can be shown as
\begin{align}
W(\textbf{x}_0,\textbf{x}_{\tau})&\leq\frac{1}{2}\int_0^{\tau}dt\sum_{j\neq i}c_{ij}|\phi_{ij}(t)| \leq \sum_{i \neq j} J_{i j} c_{i j} \int_0^{\tau} x_i (t) d t \ = \int_0^{\tau} d t \sum_i x_i (t) \sum_{l = 1}^{\infty} \sum_{j \in (i
  [l + 1] \setminus i [l])} \frac{J}{\| i - j \|^{\alpha -
\alpha_{\varepsilon}}} \nonumber\\
  & \leq \int_0^{\tau} d t \sum_i x_i (t) \sum_{l = 1}^{\infty}
  \frac{J}{l^{\alpha - \alpha_{\varepsilon} - D + 1}} = J \varphi \zeta (\alpha - \alpha_{\varepsilon} - D + 1) \tau.
\end{align}
Hence,
\begin{equation}
    \tau\geq\frac{\mu}{J \varphi \zeta (\alpha - \alpha_{\varepsilon} - D + 1)}d_{XY}^{\alpha_{\varepsilon}}\eqqcolon\kappa_1^{\varepsilon}d_{XY}^{\alpha_{\varepsilon}},
\end{equation}
which is nothing but Eq.~(5). Then, we move to prove the inequality ${\left\langle {{P_{{n_Y} \ge {N_0} + \Delta {N_0}}}} \right\rangle _{{\rho _\tau }}} \le \wp$ 
where
\begin{equation}
   \wp  = {\left( {\Delta {N_0}d_{XY}^{{\alpha _\varepsilon }}} \right)^{ - 1}}NJ\varphi \zeta (\alpha  - {\alpha _\varepsilon } - D + 1)\tau. 
\end{equation}
with the definition of the projection $P_{n_X\geq N_0}$ as
\begin{equation}
    P_{n_X\geq N_0}:=\sum_{\vec{N}:\langle\vec{N}|\hat{n}_X|\vec{N}\rangle\geq N_0}\Pi_{\vec{N}}.
\end{equation}
Therefore, $\left\langle {{P_{n_Y \ge {N_0} + \Delta {N_0}}}} \right\rangle _{{\rho _\tau }}$ is the expectation value of the projection operator and represents the probability of observing equal to or more than $N_0+\Delta N_0$ in the region $Y$ at time $\tau$. For convenience, we also represent it as $P_{n_Y(\tau) \ge {N_0} + \Delta {N_0}}$ below and in the main text.
By defining the probability distribution $p_{\vec{N}} (t) \coloneqq \mathrm{tr} (\Pi_{\vec{N}} \rho_t)$, where ${\Pi _{\vec N}} = |\vec N\rangle \langle \vec N|$ be the projection onto the state $|\vec N\rangle $, the time evolution for $p_{\vec{N}} (t)$ can be derived from the von Neumann equation as
\begin{eqnarray}
{{\dot p}_{\vec N}}(t) &=&  - i{\rm{tr}}\left( {{\Pi _{\vec N}}\left[ {{H_t},{\rho _t}} \right]} \right) = i\sum\limits_{i \ne j} {\langle {\vec N} |[ {{\rho _t},{J_{ij}}(t)b_i^\dag {b_j}} ]| {\vec N} \rangle } = i\sum\limits_{i \ne j} {{J_{ij}}} (t)\sqrt {{n_i}{n_{j'}}} ( {\langle {\vec N} |{\rho _t}| {\vec N'} \rangle  - \langle {\vec N'} |{\rho _t}|\vec N\rangle } ) \nonumber\\
 &\eqqcolon& \sum\limits_{i \ne j} {{\varphi _{\vec N\vec N'}}} \left( t \right), \label{AII-probability distribution}
\end{eqnarray}
where ${\varphi _{\vec N\vec N'}}\left( t \right)$ represents all possible flows from state ${|\vec N\rangle }$ to state ${|\vec N'\rangle }$, which satisfies $n_i' =
n_i - 1, n_j' = n_j + 1$ and $n_k' = n_k$ for all $k \neq i, j$. For the neighboring states ${|\vec N\rangle }$ and ${|\vec N'\rangle }$, the transport cost is defined as $c_{\vec{N}
\vec{N}^{'}} =\|i - j\|^{\alpha_{\varepsilon}}$. Following that, the cost between arbitrary two states ${|\vec N\rangle }$ and ${|\vec M\rangle }$ can be defined as the shortest-path cost over all possible paths connecting these states, ${c_{\vec M\vec N}} = \min \sum\nolimits_{k = 1}^K {{c_{{{\vec N}_k},{{\vec N}_{k - 1}}}}}$, where $c_{\vec{N}_0} = c_{\vec{N}}, c_{\vec{N}_K} = c_{\vec{M}}$, and ${{|\vec N}_{k - 1}\rangle}$ and ${{|\vec N}_{k}\rangle}$ are neighboring states for all $1 \le k \le K$. The Wasserstein distance reads:
\begin{equation}
  W (\textbf{p}_0, \textbf{p}_{\tau}) = \min_{\pi  \in \mathcal{C}\left( {\textbf{p}_0, \textbf{p}_{\tau}} \right)} \sum_{\vec{M}, \vec{N}}
  c_{\vec{M} \vec{N}} \pi_{\vec{M} \vec{N}},
\end{equation}
where $\textbf{p}_{\tau}$ and $\textbf{p}_0$ represent the final and initial probability distributions, respectively. Similar to the previous analysis, we obtain
\begin{equation}
  W (\textbf{p}_0, \textbf{p}_{\tau})\leq\frac{1}{2}\int_0^\tau {dt\sum\limits_{\vec N} {\sum\limits_{i \ne j} {{c_{\vec N\vec N'}}\left|\varphi _{\vec N\vec N'}(t) \right|} } }  \label{AII-Wasserstein3}.
\end{equation}
By combining Eq.~(\ref{AII-Wasserstein3}) with the definition for ${\varphi_{\vec{N} \vec{N}^{'}}}$ in Eq.~(\ref{AII-probability distribution}), the right-hand side of Eq.~(\ref{AII-Wasserstein3}) can be upper bounded as
\begin{eqnarray}
  \frac{1}{2} \int_0^\tau \sum_{\vec{N}} \sum_{i \neq j} c_{\vec{N} \vec{N}^{'}}
  | \varphi_{\vec{N} \vec{N}^{'}} | d t \leq  \sum_{\vec{N}}
  \int_0^{\tau} d t p_{\vec{N}} (t) \sum_i n_i \sum_{j (\neq i)} \frac{J}{\|i
  - j\|^{\alpha - \alpha_{\varepsilon}}}  \leq  N \tau J \varphi \zeta (\alpha - \alpha_{\varepsilon} - D + 1), \label{eq:upper} 
\end{eqnarray}
where we apply the Cauchy-Schwarz inequality in the first inequality as
\begin{equation}
|\varphi_{\vec{N}\vec{N}'}(t)|\leq|J_{ij}|[n_ip_{\vec{N}}(t)+n_j'p_{\vec{N}'}(t)].
\end{equation}
Next, we make the following definitions:
\begin{equation}
    S_0\coloneqq\{\vec{N}|\sum_{i\in X^c}n_i\leq N_0\},\quad S_{\tau}\coloneqq\{\vec{N}|\sum_{i\in Y}n_i\geq N_0+\Delta N_0\}.
\end{equation}
The process $\vec{N} \in S_0
\rightarrow \vec{M} \in S_{\tau}$ signifies that at least $\Delta N_0$ particles are transported from region $X$ to the region $Y$. To determine the lower bound of the Wasserstein distance, we first give the lower bound of $c_{\vec{M} \vec{N}}$ as
\begin{eqnarray}
  c_{\vec{M} \vec{N}} \geqslant \Delta N_0 \sum_{l = 1}^L c_{\vec{N}_l,
  \vec{N}_{l - 1}} \geqslant \Delta N_0 \sum_{l = 1}^L \|i_l - i_{l - 1}
\|^{\alpha_{\varepsilon}} \geqslant \Delta N_0 d_{X
Y}^{\alpha_{\varepsilon}}
\end{eqnarray}
with $\vec{N}_0 \rightarrow \cdots \rightarrow \vec{N}_L$ be a sequence of states that transfers one particle from $X$ to $Y$. Following this, we obtain
\begin{eqnarray}
  W (\textbf{p}_0, \textbf{p}_{\tau})  =  \min_{\pi} \sum_{\vec{M},
  \vec{N}} c_{\vec{M} \vec{N}} \pi_{\vec{M} \vec{N}}  \geqslant  \Delta N_0 d_{X Y}^{\alpha_{\varepsilon}}  \min_{\pi} \!\!
  \sum_{\vec{M} \in S_{\tau}, \vec{N} \in S_0} \!\!\!\!\!\! \pi_{\vec{M} \vec{N}},
\end{eqnarray}
and $\sum_{\vec{N} \in S_0} \pi_{\vec{M} \vec{N}} = p_{\vec{M}}
(\tau)$ since $p_{\vec{N}} (0) = 0$ for $\vec{N} \notin S_0$. Using these facts, we obtain a lower bound for the Wasserstein distance:
\begin{eqnarray}
  W (\textbf{p}_0, \textbf{p}_{\tau}) \geqslant \Delta N_0 d_{X
Y}^{\alpha_{\varepsilon}} \sum_{\vec{M} \in S_{\tau}} p_{\vec{M}} (\tau) =
  \Delta N_0 d_{X Y}^{\alpha_{\varepsilon}} \langle P_{n_Y \geqslant N_0 +
  \Delta N_0} \rangle_{\rho_{\tau}}.\label{eq:lower-WD}
\end{eqnarray}
Combining with Eq.~(\ref{eq:upper}) and Eq.~(\ref{eq:lower-WD}) yields
\begin{equation}
  P_{n_Y(\tau) \geqslant N_0 + \Delta N_0}  \leq
  \frac{N J \varphi \zeta (\alpha - \alpha_{\varepsilon} - D + 1) \tau}{\Delta
  N_0 d_{X Y}^{\alpha_{\varepsilon}}},
\end{equation}
which is the same as Eq.~(17) in the main text by assuming $\Delta N_0=\mu N$, $N_0=x_{X^c}(0)N$ and $n_{Y}(\tau)=x_{Y}(\tau)N$.

\section{Triangle inequality of the generalized Wasserstein distance} \label{Appendix-B}
In this section, we aim to prove that the generalized Wasserstein distance defined in Eq.~(20) satisfies the triangle inequality:
\begin{equation}
    \tilde{W}(\textbf{x},\textbf{y})+\tilde{W}(\textbf{y},\textbf{z})\geq\tilde{W}(\textbf{x},\textbf{z}),
\end{equation}
where $\|\textbf{x}\|_1\geq\|\textbf{y}\|_1\geq\|\textbf{z}\|_1$.\\

\textit{Proof.}~We define $\textbf{x}^{(1)}$ and $\textbf{y}^{(1)}$ be the optimal vectors satisfying $\|\textbf{x}^{(1)}\|_1=\|\textbf{y}^{(1)}\|_1$ to attain the generalized Wasserstein distance $\tilde{W}(\textbf{x},\textbf{y})$, i.e., $W(\textbf{x}^{(1)},\textbf{y}^{(1)})=\tilde{W}(\textbf{x},\textbf{y})$ where $W(\textbf{x}^{(1)},\textbf{y}^{(1)})$ is defined as
    \begin{equation}
W(\textbf{x}^{(1)},\textbf{y}^{(1)})\coloneqq\min_{\pi}\sum_{m,n}\pi_{mn}^{(1)}c_{mn}.
    \end{equation}
    Similarly, we can also define the vectors $\textbf{y}^{(2)}$ and $\textbf{z}^{(2)}$ to attain the generalized Wasserstein distance $\tilde{W}(\textbf{y},\textbf{z})$ with $\|\textbf{y}^{(2)}\|_1=\|\textbf{z}^{(2)}\|_1$. Similarly, we also have
\begin{equation}
W(\textbf{y}^{(2)},\textbf{z}^{(2)})\coloneqq\min_{\pi}\sum_{m,n}\pi_{mn}^{(2)}c_{mn}.
    \end{equation}
    By defining $\pi_{mkn}\coloneqq\pi_{kn}^{(1)}\pi_{mk}^{(2)}/y_k^{(1)}$, we can verify that
    \begin{align}
\sum_n\pi_{mkn}=\sum_n\frac{\pi_{kn}^{(1)}\pi_{mk}^{(2)}}{y_k^{(1)}}=\pi_{mk}^{(2)},\quad
\sum_m\pi_{mkn}=\pi_{kn}^{(1)}\frac{y_k^{(2)}}{y_k^{(1)}}\leq\pi_{kn}^{(1)}.\end{align}
By applying these two relations, we obtain
\begin{align}
    \tilde{W}(\textbf{x},\textbf{y})+\tilde{W}(\textbf{y},\textbf{z})&=W(\textbf{x}^{(1)},\textbf{y}^{(1)})+W(\textbf{y}^{(2)},\textbf{z}^{(2)})
\nonumber\\&=\sum_{k,n}\pi_{kn}^{(1)}c_{kn}+\sum_{m,k}\pi_{mk}^{(2)}c_{mk}\geq\sum_{m,n,k}\pi_{mkn}c_{mn} =\sum_{m,n}\tilde{\pi}_{mn}c_{mn},
\end{align}
where we apply the triangle inequality $c_{mk}+c_{kn}\geq c_{mn}$ and define a new coupling $\tilde{\pi}_{mn}\coloneqq\sum_k\pi_{mkn}$. The edge distributions of $\tilde{\pi}_{mn}$ are given by
\begin{align}
\sum_n\tilde{\pi}_{mn}=\sum_{n,k}\pi_{mkn}=\sum_k\pi_{mk}^{(2)}=z_m^{(2)},\quad
\sum_m\tilde{\pi}_{mn}=\sum_{m,k}\pi_{mkn}=\sum_k\pi_{kn}^{(1)}\frac{y_k^{(2)}}{y_k^{(1)}}\eqqcolon\tilde{x}_n,
\end{align}
where we have $\tilde{x}_n\leq x_n^{(1)}\leq x_n$, indicating $\textbf{x}\succeq\tilde{\textbf{x}}$. Consequently, we derive the following inequality:
\begin{align}
\tilde W(\textbf{x},\textbf{y}) + \tilde W(\textbf{y},\textbf{z}) \ge \sum_{mn} {{{\tilde \pi }_{mn}}{c_{mn}}}\ge W(\tilde {\textbf{x}},{{\textbf{z}}^{(2)}}) \ge \tilde W(\textbf{x},\textbf{z}),
\end{align}
which establishes the triangle inequality for the generalized Wasserstein distance and completes the proof. $\square$

\end{document}